\documentclass[twocolumn,prb,amssymb,showpacs]{revtex4}
\usepackage{bm}

\usepackage{graphicx}

\begin{document}
\title{
Weak antilocalization in quantum wells in tilted magnetic fields}

\author{G.M.~Minkov}
\author{A.V.~Germanenko}
\author{O.E.~Rut}
\author{A.A.~Sherstobitov}

\affiliation{Institute of Physics and Applied Mathematics, Ural
State University, 620083 Ekaterinburg, Russia}

\author{L.E.~Golub}
\affiliation{A.F.~Ioffe Physico-Technical Institute, Russian
Academy of Sciences, 194021 St.~Petersburg, Russia}

\author{B.N.~Zvonkov}
\affiliation{Physical-Technical Research Institute, University of
Nizhni Novgorod, 603600 Nizhni Novgorod, Russia}

\author{M.~Willander}
\affiliation{Physical Electronics and Photonics, Department of
Physics, Chalmers University of Technology and G\"{o}teborg
University, S-412~96, G\"{o}teborg, Sweden}

\date{\today}
\begin{abstract}
Weak antilocalization is studied in an InGaAs quantum well.
Anomalous magnetoresistance is measured and described
theoretically in fields perpendicular, tilted and parallel to the
quantum well plane. Spin and phase relaxation times are found as
functions of temperature and parallel field. It is demonstrated
that spin dephasing is due to the Dresselhaus spin-orbit
interaction. The values of electron spin splittings and spin
relaxation times are found in the wide range of 2D density.
Application of in-plane field is shown to destroy weak
antilocalization due to competition of Zeeman and microroughness
effects. Their relative contributions are separated, and the
values of the in-plane electron $g$-factor and characteristic size
of interface imperfections are found.
\end{abstract}


\maketitle

\section{Introduction}

The spin properties of carriers have attracted much attention in
recent years due to rapidly developed spintronics dealing with the
manipulation of spin in electronic
devices.~\cite{spintronicbook02} Particular attention is paid to
semiconductor quantum wells (QWs) and other heterostructures,
whose spin properties can be controlled by advanced technology. A
very powerful method of study of the spin properties are
magnetotransport investigations. Starting with the very first
experiments~\cite{SdHh,SdHe} entailing theoretical
explanations~\cite{Bychkov84p78} and up to now, spin-related
phenomena have been considering in a large number of papers.
Performing studies in magnetic fields tilted in respect to the QW
plane opens new possibilities in comparison to bulk structures
because the perpendicular field component, $B_\perp$, affects
mostly orbital motion while the in-plane magnetic field,
$B_\parallel$, couples only with the carrier spins.

One of the magnetotransport tools for study spin effects is known
to be measurements of anomalous low-field magnetoresistance. It is
caused by \textit{weak localization} phenomenon that consists in
interference of paths passing by a scattered particle forward and
back, i.e. of time-reversed paths.~\cite{AA} It has been
established that particles with spin behave differently: the
anomalous magnetoresistance is caused by competition of weak
localization with so-called \textit{weak
antilocalization}.~\cite{HLN} Spin relaxation processes change its
sign from negative to positive that has been measured in
three-dimensional (3D) systems and metal films.~\cite{Bergmann}

In semiconductor QWs, positive magnetoresistance measurements in a
field perpendicular to the 2D structure has been first reported a
decade ago.~\cite{Dress,InAsWAL} However attempts to describe the
data by the old theory~\cite{HLN} failed. The reason was in the
nature of spin-orbit interaction in semiconductor
heterostructures. It has the form of odd in the 2D wavevector
spin-dependent terms in the Hamiltonian that are caused by
structure inversion asymmetry (the Rashba term) and by the bulk or
interface inversion asymmetry (the Dresselhaus term). A new theory
taking into account both Rashba and Dresselhaus spin-orbit
interactions has been developed in Ref.~\onlinecite{ILP}. It has
been shown that effect of the spin-orbit interaction on weak
localization does not reduced to only spin relaxation but also
manifests itself as a spin precession in the effective magnetic
field caused by Rashba or Dresselhaus terms. The correct
expressions for the anomalous magnetoresistance have been derived.
This allowed to describe successfully the experimental data of
Ref.~\onlinecite{Dress} and reliably extract the spin-orbit,
spin-relaxation and dephasing parameters.~\cite{PikusPikus} Soon
afterwards the new experiments have been performed and the results
occurred to be in excellent agreement with the new
theory.~\cite{Knap,110} Similar studies have been performed during
all the recent decade allowing to investigate the carrier spin
properties in 2D systems hitherto.~\cite{WAL&SIA}

In a magnetic field parallel to QW interfaces, the anomalous
magnetoresistance also takes place due to weak localization
effects. It has been shown theoretically that the Zeeman
interaction destructs weak antilocalization that results in
suppressing positive magnetoresistance.~\cite{Malsh1,Malsh2}
Another effect of $B_\parallel$ is due to microroughness present
in QW structures.~\cite{Malsh1,Malsh2,Rough} The parallel field
leads to an additional dephasing that also reveals itself in
negative magnetoresistance. Experiments on weak localization in
parallel fields have been performed however only on quantum
dots.\cite{Dots}

In this paper we investigate effects of spin-orbit and Zeeman
interactions on weak antilocalization in semiconductor QWs. We
perform anomalous magnetoresistance measurements in magnetic
fields tilted in respect to the 2D plane. We show that weak
antilocalization is suppressed by $B_\parallel$. The effects of
Zeeman splitting and microroughness on anomalous magnetoresistance
are separated.

\section{Experimental details}
\begin{figure}
\includegraphics[width=\linewidth,clip=true]{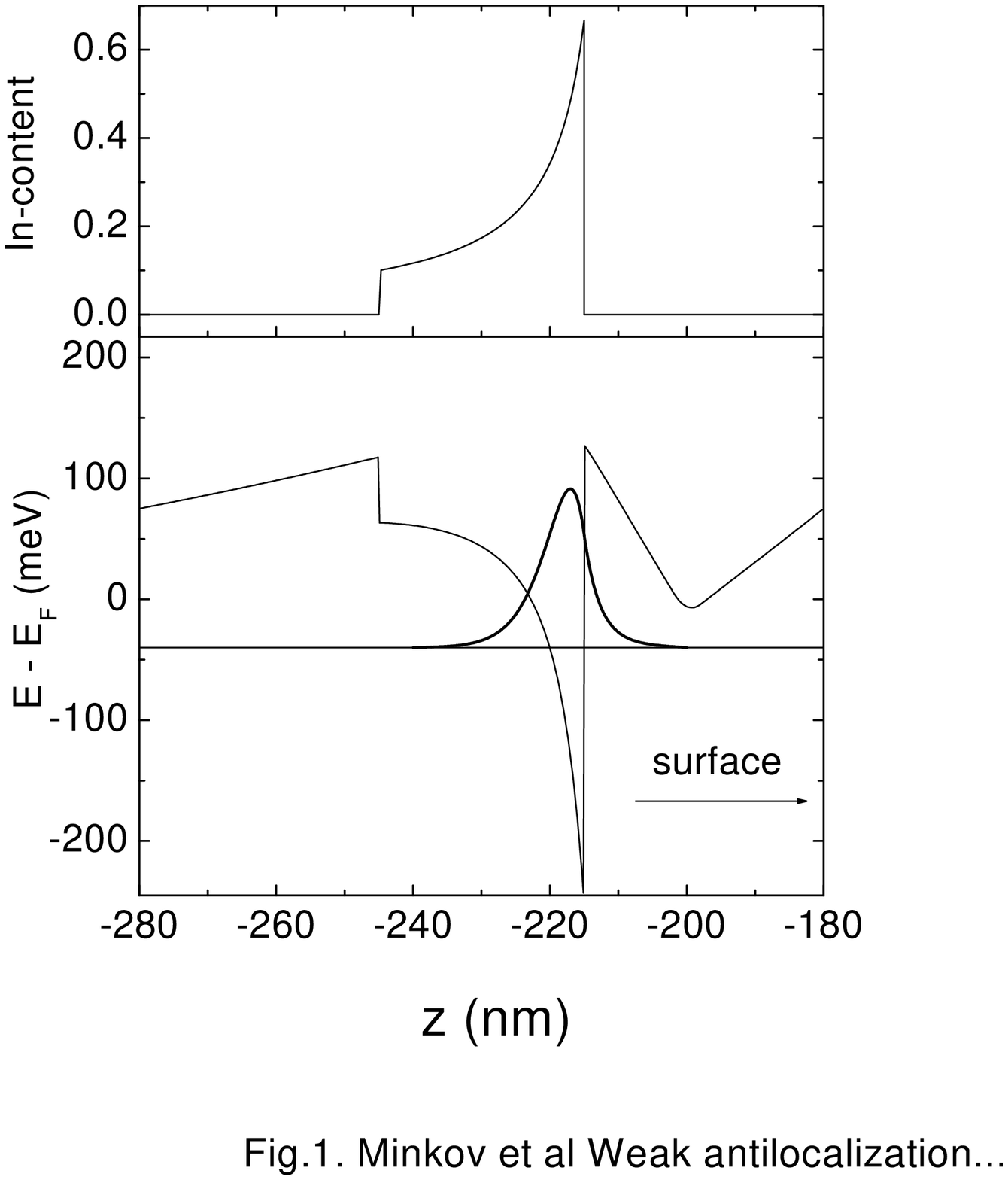}
\caption{Calculated energy diagram and squared wave function for
the investigated structure.}
 \label{f1}
\end{figure}

The quantum well GaAs/In$_{x}$Ga$_{1-x}$As/GaAs heterostructure
was grown by metal-organic vapor phase epitaxy on semi-insulator
GaAs substrate. It consists of a 0.3~$\mu$m-thick undoped GaAs
buffer layer, a 30~nm In$_{x}$Ga$_{1-x}$As QW, a 15~nm spacer of
undoped GaAs, a Si $\delta$-layer, and  200~nm cap layer of
undoped GaAs. The concentration of In within the QW varies from
$0.1$ to $0.6$ from the buffer to cap as $0.6/[6-0.17(z+245)]$,
where $z$ is the coordinate perpendicular to the quantum well
plane, measured in nanometers [upper panel in Fig.~\ref{f1}]. The
energy diagram calculated self-consistently\cite{slfcn} for this
structure is presented in the lower panel in Fig.~\ref{f1}. The
electron density $n$ and mobility $\mu$ are the following:
$n=8\times 10^{15}$~m$^{-2}$, $\mu=2.4$~m$^2$/(Vs).

The samples were mesa etched into  Hall bars with tree potential
probes along each side and then an Ag or Al gate electrode was
deposited by thermal evaporation onto the cap layer through a
mask. The width of the bars and  distance between the potential
probes were 0.5~mm and 1~mm, respectively. Although varying the
gate voltage $V_g$ we were able to change the density of electron
gas in the QW from $4\times 10^{15}$ to $8.5\times
10^{15}$~m$^{-2}$, the most part of results presented here relate
to  $V_g=0$ (exception is Fig.~\ref{f5}). The parameters of
electron gas for $V_g=0$ are the following: $n=8\times
10^{15}$~m$^{-2}$, $\sigma\simeq 3.1\times
10^{-3}$~$\Omega^{-1}\Box$ at $T=1.4$~K, the mean free path
$l=370$~nm, the effective electron mass obtained from the
Shubnikov-de Haas experiment is $(0.053\pm 0.005)m_0$, the
diffusion coefficient $D$ is approximately equal to
$0.09$~m$^2$/s.

\begin{figure}
\includegraphics[width=\linewidth,clip=true]{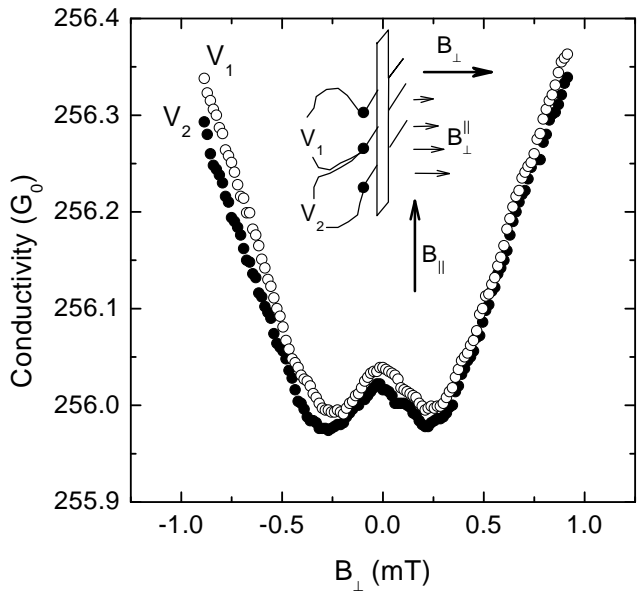}
\caption{The conductivity as a function of perpendicular magnetic
field measured at $B_\parallel=0.3$~T and $T=1.4$~K for two
neighbor pairs of potential probes as shown in inset. Arrows in
inset marked as  $B_\perp^\parallel$ show perpendicular component
of in-plane magnetic field which varies along the sample due to
imperfection of magnet system.} \label{f2}
\end{figure}

The transverse magnetoresistance was measured in perpendicular
magnetic field up to 2~mT with step $0.01$~mT within the
temperature range $0.45-5.0$~K. In order to apply tesla-scale
in-plane magnetic field while sweeping subgauss control of
perpendicular field, we mount the sample with 2D electrons aligned
to the axis of primary solenoid (accurate to $\sim 1^\circ$) and
use an independent split-coil solenoid  to provide $B_\perp$ as
well as to compensate for sample misalignment. The two calibrated
Hall probes were used to measure $B_\perp$ and $B_\parallel$.
Since antilocalization behavior of the magnetoresistance is
observed at perpendicular field less than $0.2$~mT it is very
important that the gradient of the perpendicular component of the
tesla-scale in-plane magnetic field $B_\perp^\parallel$ was  less
than 0.01 mT/mm. To assure that this condition is fulfilled the
magnetoresistance was measured simultaneously on two neighbor
potential pairs placed along one side of the bar (see inset in
Fig.~\ref{f2}).
The shift of these magnetoresistance curves relative to each other
in $B_\perp$-direction is proportional to the gradient. To
decrease the gradient along the structure there was needed to
remove all, even slight, magnetic details and find an optimal
position of the sample in primary solenoid. Fig.~\ref{f2}  shows
that as a result of such procedures we have reduced the gradient
of perpendicular component of in-plane magnetic field down to
$1.6~10^{-4} B_{\parallel}/$mm.

\section{Magnetoresistance in a perpendicular field}
\label{sec:perp}

Let us discuss the low-field magnetoresistance at zero in-plane
field. The expression for the magnetoconductivity in a
perpendicular field has the form
\begin{equation}\label{dsigma_perp}
  \sigma(B_{\perp}) - \sigma(0) =  {G_0 \over 2} \left[ F_t \left( b_\phi,
  b_s
  \right) - F_s \left( b_\phi \right)\right],
\end{equation}
where $G_0 = e^2 / 2 \pi^2 \hbar$, $b_\phi=B_\phi/B_{\perp}$, and
$b_s=B_s/B_{\perp}$. The first term in Eq.~(\ref{dsigma_perp}) is
the interference contribution of scattered electrons with the
total momentum equal to one, i.e. being in the triplet state, and
the second term is the singlet part. The latter depends only on
one parameter $B_\phi = \hbar / 4 e D \tau_\phi$, where
$\tau_\phi$ is the phase relaxation time. The triplet contribution
depends not only on the dephasing rate but also on the spin
relaxation time $\tau_s$ via the parameter $B_s = \hbar / 4 e D
\tau_s$.

Spin relaxation in InGaAs QWs with low In content is caused by
spin splitting of electron spectrum (the D'yakonov-Perel'
mechanism). The Hamiltonian of corresponding spin-orbit
interaction is given by
\begin{equation}\label{H_SO}
  H_{SO}( \bm{k}) = \hbar {\bm \sigma} \cdot {\bm \Omega}({\bm k}),
\end{equation}
where $\bm{\sigma}$ is the vector of Pauli matrices and the 2D
vector in the plane of QW, $\bm{\Omega}$, is an odd function of
${\bm k}$. There are the Rashba ($\bm{\Omega}_R$) and Dresselhaus
($\bm{\Omega}_D$, $\bm{\Omega}_3$) contributions to $\bm \Omega$.
The Rashba term for Fermi electrons has the form
\begin{equation}\label{OmegaR}
    \hbar \bm{\Omega}_R = \alpha k_{\rm F} (\sin{\varphi}, -\cos{\varphi}),
\end{equation}
where $\varphi$ is the angle between $\bm k$ and the [100] axis,
$k_{\rm F}$ is the Fermi wavevector, and the constant $\alpha$ is
given by\cite{alpha1,alpha2,alpha3}
\begin{equation}
\alpha= {P^2 \over 3} \int dz \, \psi^* {d\over dz}\left[ {1 \over
E_{\rm F}-E_{\Gamma_7}(z)}-{1 \over E_{\rm F}
-E_{\Gamma_8}(z)}\right]\psi.\label{eq3}
\end{equation}
Here, $\psi$ is the wave function of 2D electrons, $P$ is the Kane
matrix element, $E_{\rm F}$ is the Fermi energy, and
$E_{\Gamma_7}(z)$ and $E_{\Gamma_8}(z)$ are the band edge energies
for $\Gamma_7$ and $\Gamma_8$ valence bands, respectively, at
position $z$.

The Dresselhaus term $\bm{\Omega}_D$ also being the first
Fourier-harmonic of $\varphi$, has the following form in
[001]-grown QWs
\begin{equation}\label{OmegaD}
    \hbar \bm{\Omega}_D = \gamma k_{\rm F} \left(\langle k_z^2 \rangle - {k_{\rm F}^2 \over 4}\right) (\cos{\varphi}, -\sin{\varphi}),
\end{equation}
where $\gamma$ is the bulk constant of spin-orbit interaction, and
$\langle k_z^2 \rangle$ is the mean square of electron momentum in
the growth direction
\begin{equation}
\langle k_z^2 \rangle = \int dz \, \psi^*  \left( - {d^2 \over
dz^2} \right) \psi.
\end{equation}

The third harmonic of the Dresselhaus term, $\bm{\Omega}_3$, is
given by
\begin{equation}\label{Omega3}
    \hbar \bm{\Omega}_{3} =  \gamma {k_{\rm F}^3 \over 4} \: (\cos{3 \varphi}, -\sin{3 \varphi}).
\end{equation}

The theory of interference induced magnetoresistance in 2D systems
with spin-orbit interaction Eq.~(\ref{H_SO}) has been developed in
Ref.~\onlinecite{ILP}. The authors derived an analytical
expression for the magnetic field dependence of the conductivity
when only $\Omega_D$ and $\Omega_3$ (or $\Omega_R$ and $\Omega_3$)
contribute to the interference correction. Below we show that the
magnetoresistance curves for the studied sample are well described
with taking into account only $\Omega_D$. In this case, the spin
relaxation time for the parameter $B_s$ is given by the expression
\begin{equation}\label{tau_s}
  {1 \over \tau_s} = 2 \Omega_D^2 \tau_p,
\end{equation}
where $\tau_p$ is the momentum scattering time. It is worth to
mention that $\tau_s$ in the above expression coincides with the
D'yakonov-Perel' relaxation time for a spin lying in the QW plane.

The expressions for the functions $F_s$ and $F_t$ are derived in
Ref.~\onlinecite{ILP} in a so-called diffusion approximation. It
is valid for magnetic fields $B_\perp \ll B_{tr}$, where $B_{tr} =
\hbar/4 e D \tau_p$ is the ``transport'' field. A theory for high
fields is developed in Ref.~\onlinecite{WAL_Yulik_PRL} accounting
for the presence of all three terms in $\bm \Omega$. The obtained
expressions, however, are valid for $B_\perp \gg B_s$. In our
samples with moderate mobility, the transport field is strong
enough ($B_{tr}=2.4$~T) so that the antilocalization minimum in
magnetoconductivity takes place at $B_\perp \sim B_s < B_{tr}$.
This allows us to extract the dephasing times $\tau_s$,
$\tau_\phi$ from the low-field range where the diffusion
approximation is valid.

Under these circumstances, the function $F_s(b_\phi)$ is given by
\begin{equation}\label{Fs}
  F_s(b_\phi) = \Psi(1/2 + b_\phi) - \ln{b_\phi},
\end{equation}
where $\Psi$ is the digamma-function. The expression for $F_t(
b_\phi, b_s )$ is as follows\cite{ILP,Lenya}
\begin{widetext}
\begin{eqnarray}\label{Ft}
F_t( b_\phi, b_s ) &=& \sum_{n=1}^\infty \left\{ {3 \over n} - {3
a_n^2 + 2 a_n b_s - 1 - 2(2n+1) b_s \over (a_n+b_s)a_{n-1}a_{n+1}
- 2 b_s [(2n+1)a_n - 1]} \right\} - {1 \over a_0} - {2 a_0 + 1 +
b_s \over a_1(a_0+b_s) - 2 {b_s}} \nonumber \\
&-& 2 \ln{(b_\phi + b_s)} - \ln{(b_\phi + 2 b_s)} - 3C -
S(b_\phi/b_s)
\end{eqnarray}
\end{widetext}
with $C \approx 0.57721$ as the Euler constant, $a_n = n + 1/2 +
b_\phi + b_s$, and the $B_\perp$-independent function
$S(b_\phi/b_s)$ is given by
\begin{equation}\label{S}
    S(x)={8 \over \sqrt{ 7 + 16x}} \left[\arctan{ \left( { \sqrt{ 7 + 16x} \over 1 - 2
    x} \right)}-\pi\Theta(1-2x)\right],
\end{equation}
where $\Theta(y)$ is the Heaviside step function.

\begin{figure}
\includegraphics[width=\linewidth,clip=true]{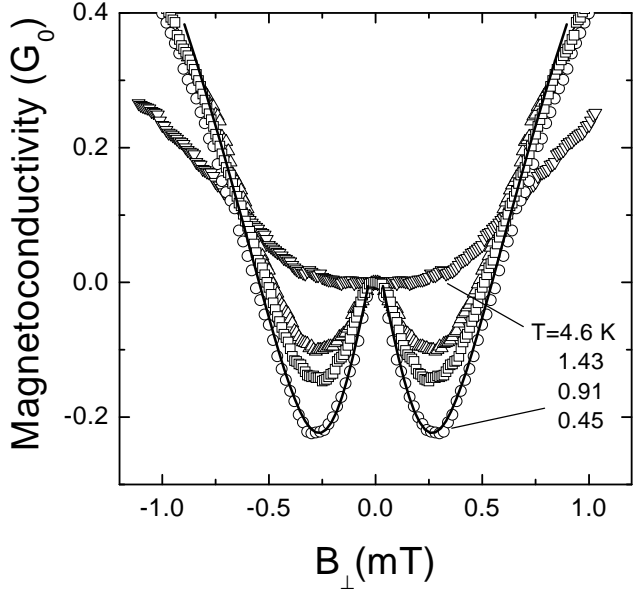}
\caption{The low-field magnetoconductivity at different
temperatures for $B_\parallel=0$ (symbols). Solid line is the best
fit by Eq.~(\ref{dsigma_perp})  with the parameters
$\tau_\phi=72$~ps and $\tau_s=8.8$~ps.} \label{f3}
\end{figure}

\begin{figure}
\includegraphics[width=\linewidth,clip=true]{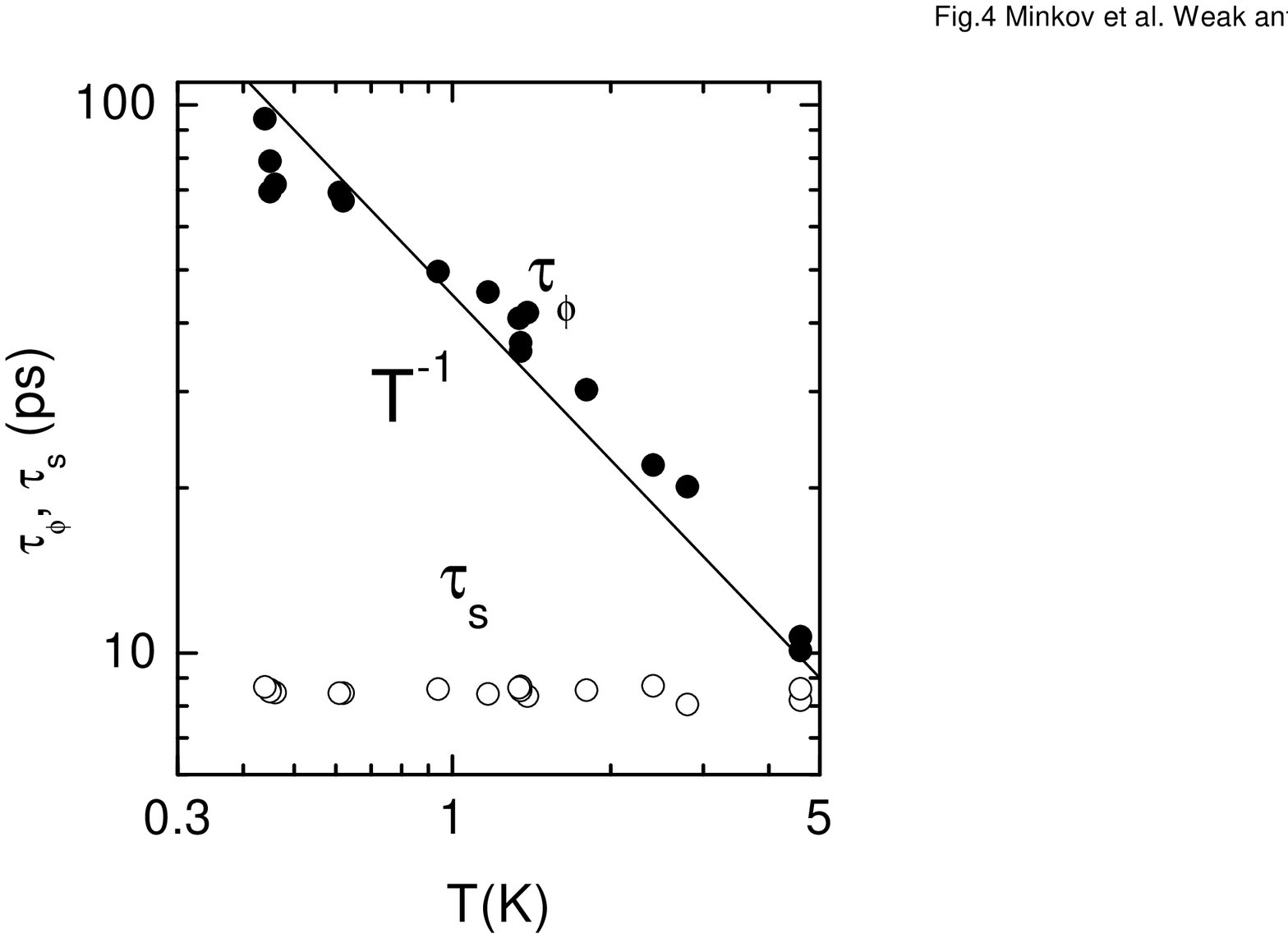}
\caption{The temperature dependences of the phase and spin
relaxation times for $V_g=0$ and $B_\parallel=0$.} \label{f4}
\end{figure}

The experimental curves
$\Delta\sigma(B_{\perp})=\rho_{xx}^{-1}(B_{\perp})-\rho_{xx}^{-1}(0)$
measured at $V_g=0$ for different temperatures and the fit by
Eq.~(\ref{dsigma_perp}) with $\tau_{\phi}$ and $\tau_s$  as
fitting parameters are presented in Fig.~\ref{f3}.
The fit was carried out within magnetic field range $0< B_\perp
<0.3\, B_{tr}$. The temperature dependences of the phase and spin
relaxation times found by this way are presented in Fig.~\ref{f4}.
It is evident that the behavior of $\tau_{\phi}$ is close to
$T^{-1}$-law predicted theoretically,~\cite{AA} whereas $\tau_{s}$
is temperature independent that corresponds to the
D'yakonov-Perel' mechanism of spin relaxation in a degenerate
electron gas.

\begin{figure}
\includegraphics[width=\linewidth,clip=true]{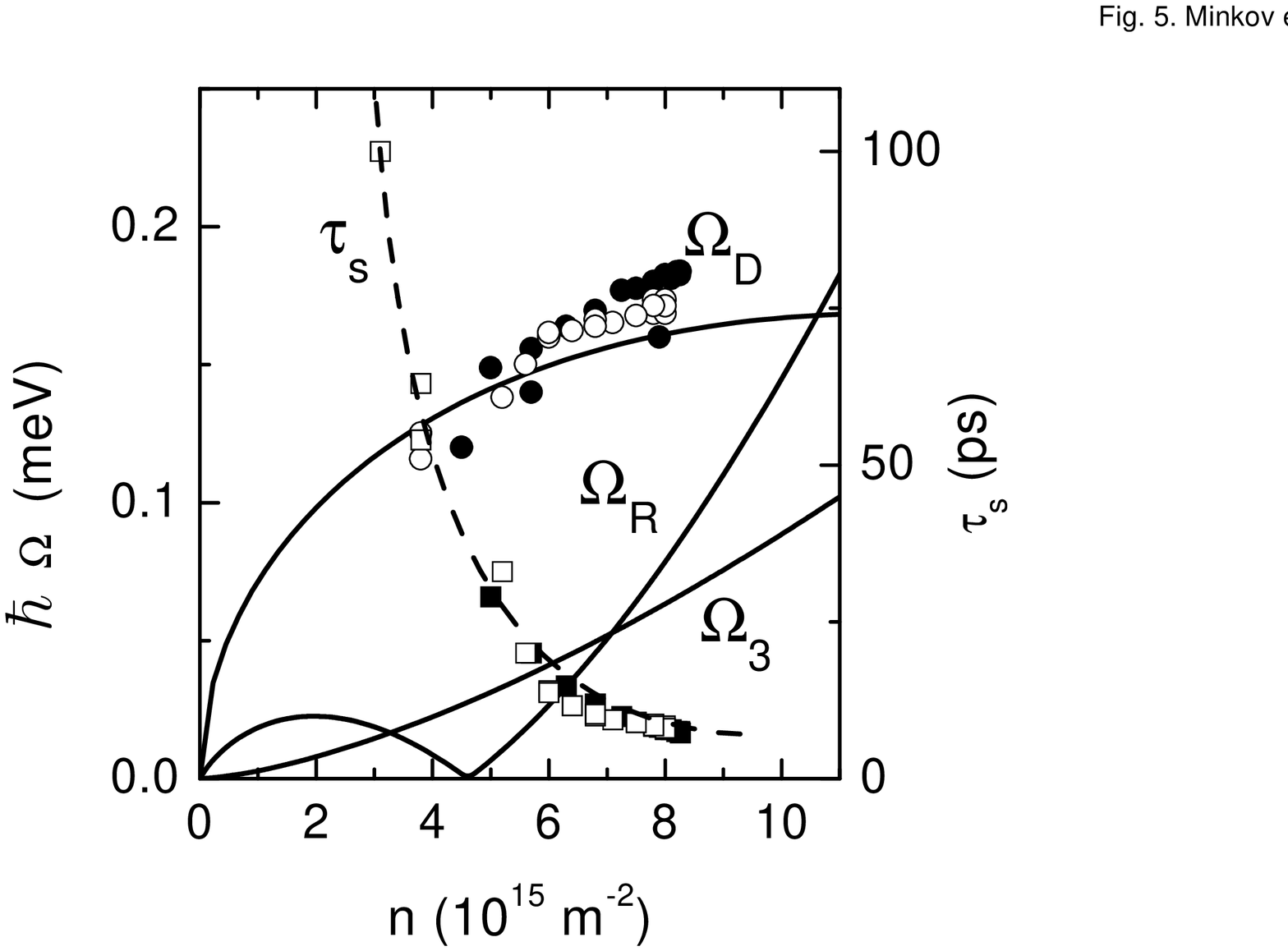}
\caption{The values of $\hbar\Omega_R$, $\hbar\Omega_D$,
$\hbar\Omega_3$ (circles) and $\tau_s$ (squares) as functions of
electron density for the investigated structures. Open and solid
symbols are the experimental data for $T=0.45$ K and $1.5$~K,
respectively. The lines are the result of self-consistent
calculations\cite{Our1} in which we used $\gamma=18$~eV\,\AA$^3$.}
\label{f5}
\end{figure}

The good agreement allows us to find the value of spin-orbit
splitting $\hbar\Omega$. Such analysis has been carried out for
wide range of electron density and final results are presented in
Fig.~\ref{f5}.
In the same figure, the electron density dependence is shown for
all three terms $\Omega_D$, $\Omega_R$, and $\Omega_3$ found from
self-consistent calculations for the studied structure by using
Eqs.~(\ref{OmegaR})-(\ref{Omega3}). One can see that the
experimental data are close to the Dresselhaus term $\Omega_D$
which dominates both  $\Omega_3$ and the Rashba term $\Omega_R$
within actual range of electron density.

\section{Transverse magnetoresistance in the presence of in-plane magnetic field}
\label{sec:tilt}

\begin{figure}
\includegraphics[width=\linewidth,clip=true]{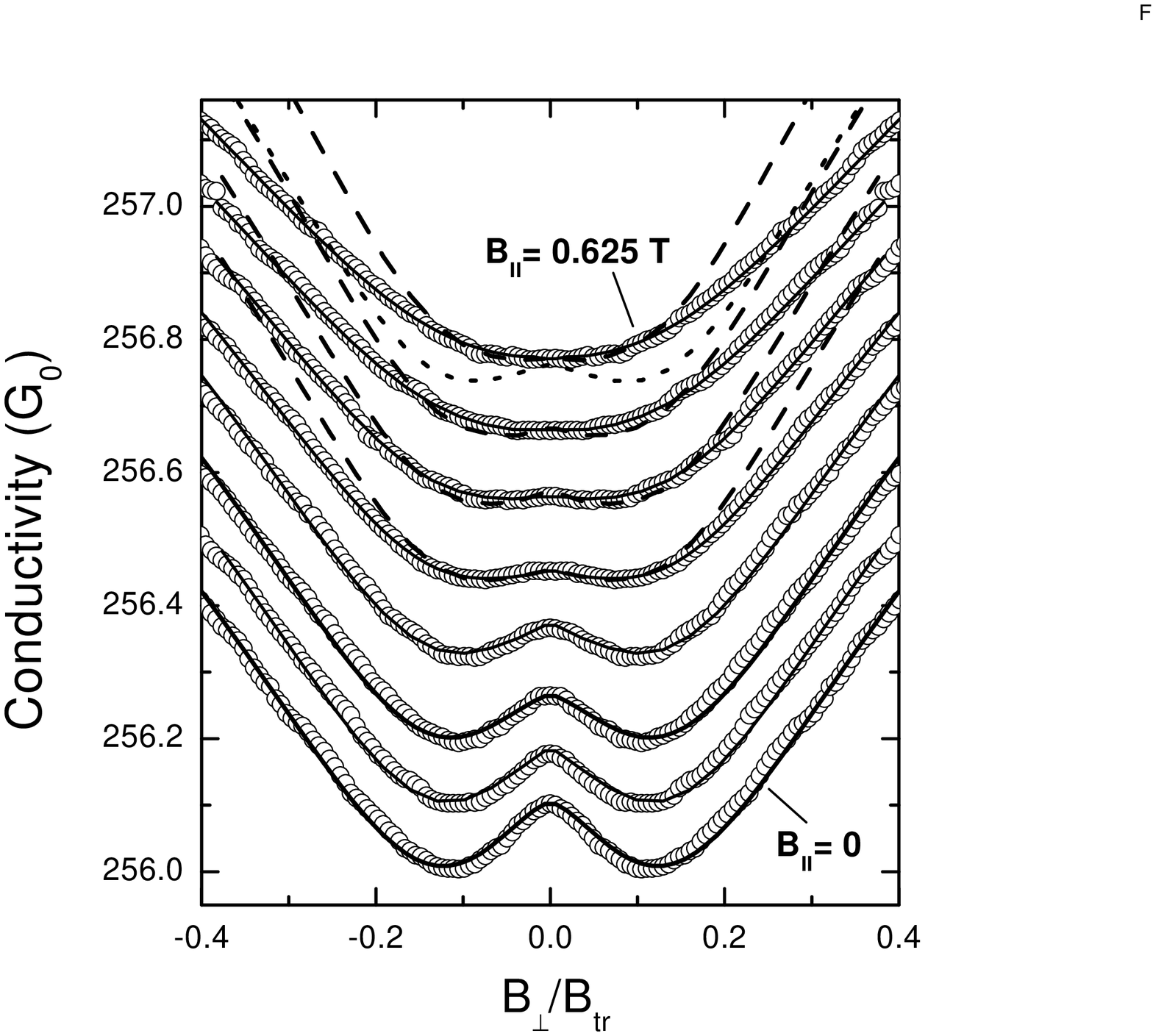}
\caption{The conductivity as a function of $B_\perp$, measured at
$T=1.4$~K for different in-plane magnetic fields $B_\parallel$:
$0$, $0.075$, $0.125$, $0.225$, $0.325$, $0.425$, $0.525$ and
$0.625$~T. The dashed curves are the best fit with only Zeeman
contribution for some values of $B_\parallel$. Dotted curve
illustrates the result of the fit with only Zeeman contribution
carried out in the range of $B_\perp/B_{tr}$ from $0$ to $0.3$.
The solid curves are the best fit when both roughness and Zeeman
splitting are taken into account carried out with
$B_\phi=0.0625$~mT and $B_s=0.28$~mT found at $B_\parallel=0$ and
$\Delta_s$ and $\Delta_r$ presented in Fig.~\ref{f7}. For clarity,
the plots are separated in vertical direction by the value of
0.1$G_0$.} \label{f6}
\end{figure}

Now let us discuss the effect of an in-plane magnetic field. The
magnetoconductance versus perpendicular magnetic field measured
for different $B_\parallel$ is presented in Fig.~\ref{f6}.
It is seen that application of in-plane magnetic field decreases
the depth of antilocalization minimum which disappears at $B\simeq
(0.5-0.6)$~T.

The influence of in-plane magnetic field on the interference
induced magnetoconductance is due to Zeeman splitting and
interface microroughness. We start our consideration with Zeeman
effect. The in-plane magnetic field couples the triplet and
singlet states of interfering carriers. This makes
Eq.~(\ref{dsigma_perp}) incorrect. However the applied magnetic
fields are not too strong in our experiments: the condition
$$g \mu_{\rm B} B_\parallel < \hbar/\tau_s$$
is met up to the highest $B_\parallel = 0.625$~T. Here $g$ is the
in-plane electron $g$-factor and $\mu_{\rm B}$ is the Bohr
magneton. In such low parallel field, dephasing of the triplet
state is still determined by zero-field values of $\tau_s$ and
$\tau_\phi$. The Zeeman interaction affects only the singlet
contribution resulting in its additional
depasing.~\cite{Malsh1,Malsh2} The corresponding correction to
$B_\phi$ in the second term of Eq.~(\ref{dsigma_perp}),
$\Delta_s$, has the form
\begin{equation}\label{Delta_s}
    \Delta_s = { \tau_s \over 4 e \hbar D} \left(g \mu_{\rm B}B_\parallel\right)^2.
\end{equation}

As a result, we get the expression for the interference induced
transverse magnetoresistance in the presence of an in-plane
magnetic field is given by
\begin{equation}\label{dsigma_tilted}
     \sigma(B_{\perp},B_\parallel) - \sigma(0,B_\parallel) = {G_0 \over
    2} \biggl[ F_t \left( b_\phi, b_s \right) - F_s \left(
    b^{\prime }_\phi\right)\biggr],
\end{equation}
where
$$
b'_\phi = {B_\phi + \Delta_s(B_\parallel) \over B_{\perp}},
$$
and the functions $F_s$, $F_t$ are given by Eqs.~(\ref{tau_s}) and
(\ref{Ft}).

Results of the fit of experimental data by
Eq.~(\ref{dsigma_tilted}) with $\tau_{\phi}$,  $\tau_s$  found at
$B_\parallel =0$ and with $\Delta_s$ as fitting parameter are
presented in Fig.~\ref{f6} by dashed lines. As seen the range of
$B_\perp$ in which a good fit can be achieved, narrows rapidly
with increasing of $B_\parallel$. If the fitting range is kept
constant, it is impossible to describe the transverse
magnetoconductivity in the presence of in-plane magnetic field
even qualitatively (dotted curve in Fig.~\ref{f6}). Thus, taking
into account only Zeeman splitting does not give an agreement with
experimental data. The reason for such discrepancy is the effect
of microroughness.

Imperfections of the QW interfaces results in that an electron
moving over QW  shifts in the perpendicular direction also. This
means that carriers in real 2D systems effectively feel random
perpendicular magnetic field. This effect gives rise to the
longitudinal magnetoresistance (considered in the next Section)
and changes the transverse magnetoresistance in the presence of an
in-plane field.

The effect of short-range roughness with $L \ll l$, where $L$ is
the correlation length of the QW width fluctuations, on the shape
of magnetoresistance is that application of an in-plane magnetic
field also leads to additional dephasing. In areas with
short-range fluctuations normal to the QW plane, carriers move
transversely to the magnetic field $B_\parallel$. Their weak
localization on these fluctuations is destroyed by $B_\parallel$.
Quantitatively, in the presence of parallel magnetic field, the
phase relaxation parameter $B_\phi$ acquires an
addition~\cite{Malsh2, Rough}
\begin{equation}\label{Delta_r}
    B_\phi(B_\parallel) = B_\phi(0) + \Delta_r(B_\parallel),
\end{equation}
where
\begin{equation}\label{Delta_rB}
\Delta_r(B_\parallel)\simeq{\sqrt{\pi} \over 2}{e \over \hbar}
{d^2 L \over l} B_\parallel^2.
\end{equation}
Here, $d$ is the root-mean-square height of the fluctuations.
Note, this effect manifests itself in the both singlet and triplet
contributions in contrast to Zeeman splitting.

Thus, the final expression for $\sigma(B_{\perp},B_\parallel) -
\sigma(0,B_\parallel)$ when both mechanisms are taken into account
is
\begin{equation}\label{dsigma_tiltedRS}
     \sigma(B_{\perp},B_\parallel) - \sigma(0,B_\parallel) = {G_0 \over
    2} \biggl[ F_t \left( b_\phi, b_s \right) - F_s \left(
    \widetilde{b}_\phi\right)\biggr],
\end{equation}
where
$$
\widetilde{b}_\phi = {B_\phi + \Delta_s(B_\parallel) +
\Delta_r(B_\parallel) \over B_{\perp}}
$$
and
$$
b_\phi = {B_\phi  + \Delta_r(B_\parallel) \over B_{\perp}}.
$$
The results of the fit by Eq.~(\ref{dsigma_tiltedRS}) with
$\tau_{\phi}$,  $\tau_s$  found at $B_\parallel =0$ and with
$\Delta_s$ and $\Delta_r$ as fitting parameters are presented in
Fig.~\ref{f6} by solid lines. It is clearly seen that an excellent
agreement occurs for all values of in-plane magnetic field.

\begin{figure}
\includegraphics[width=\linewidth,clip=true]{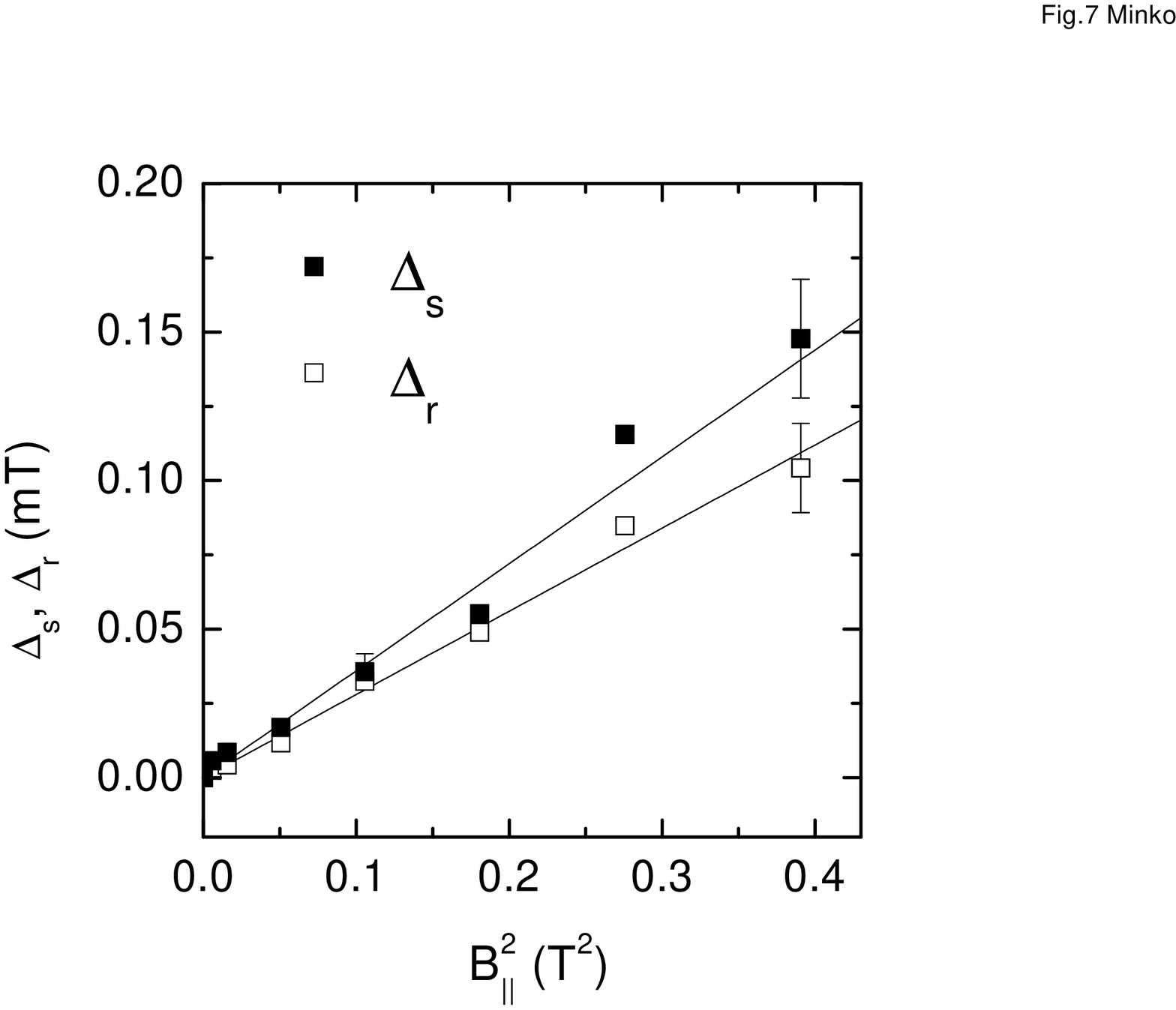}\caption{The value
of the fitting parameters $\Delta_s$ and $\Delta_r$ corresponding
to solid lines in Fig.~\ref{f6} as a function of $B^2_\parallel$
(symbols). Upper line is Eq.~(\ref{Delta_s}) with $\tau_s=8.8$~ps,
$D=0.09$~m$^2$/s and $g=1.7$, lower one is Eq.~(\ref{Delta_rB})
with $l=370$~nm and $d^2L=75$~nm$^3$.} \label{f7}
\end{figure}

The used model predicts that the parameters $\Delta_s$ and
$\Delta_r$ have to depend on $B_\parallel$ quadratically [see
Eqs.~(\ref{Delta_s}) and (\ref{Delta_rB}), respectively].
Therefore in Fig.~\ref{f7} we have presented the values of
$\Delta_s$ and $\Delta_r$ as functions of $B_\parallel ^2$. It is
seen that these dependences are really close to quadratical ones
within an experimental error. It allows us to determine the values
of in-plane $g$-factor and parameter of roughness $d^2 L$. We
obtain $g=1.7\pm0.3$ and $d^2 L=(75\pm 15)$~nm$^3$. This value of
$g$-factor corresponds to that for bulk In$_x$Ga$_{1-x}$As with
$x\simeq 0.35$,~\cite{g-factor} that in its turn agrees with
average indium content within the QW (see Fig.~\ref{f1}). The
value of $d^2 L$ occurs several times larger than that obtained in
Ref.~\onlinecite{ourProd} for analogous structures with
In$_{0.2}$Ga$_{0.8}$As quantum well. We can explain this
discrepancy by mechanical strain arising in heterostructures
In$_x$Ga$_{1-x}$As/GaAs with sufficiently high indium content due
to lattice mismatch.

Thus we successively described the anomalous magnetoresistance in
a tilted field. The effects of the in-plane field component are
shown to be important at $B_\parallel \sim 1/(g \mu_{\rm B}
\sqrt{\tau_s \tau_\phi}) \gg B_\perp$. This explains the observed
resistance independence of a weak in-plane field $B_\parallel \sim
B_\perp$.\cite{Stud}

\section{Longitudinal magnetoresistance}
\label{sec:long}

\begin{figure}
\includegraphics[width=\linewidth,clip=true]{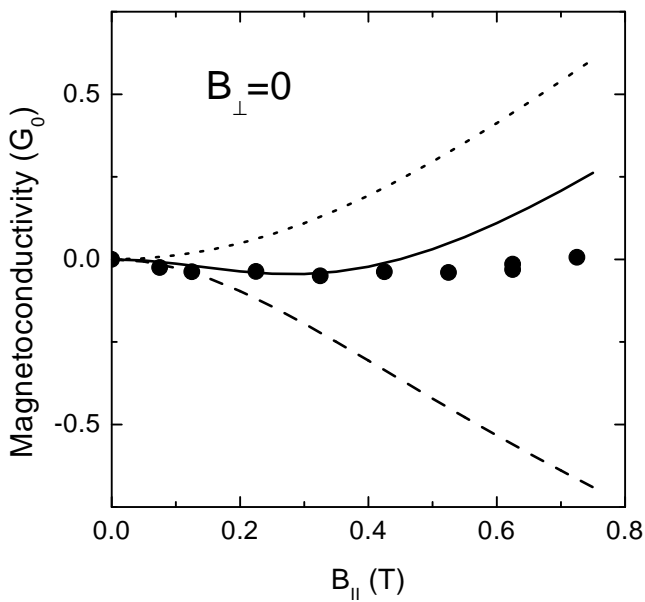}
\caption{The longitudinal magnetoconductivity  for $T=1.4$~K
(symbols). Solid line is Eq.~(\ref{inplaneMR}) with
$B_\phi=0.0625$~mT, $B_s=0.28$~mT, and $\Delta_r(B_\parallel)$ and
$\Delta_s(B_\parallel)$  shown in Fig.~\ref{f7} by solid lines.
Dotted and dashed lines are calculation results when only
roughness ($\Delta_s=0$) or Zeeman effect ($\Delta_r=0$) is taken
into account, respectively. } \label{f8}
\end{figure}

Let turn now to the effects of pure in-plane field. Longitudinal
magnetoresistance measured at T=1.4~K is shown by points in
Fig.~\ref{f8}. As mentioned above the in-plane magnetic field does
not destroy the electron interference of absolutely flat 2D gas of
spinless particle and does not result in longitudinal
magnetoresistance. However, in real systems this effect should be
evident due to roughness of interface. Its magnitude can be easily
estimated as\cite{Rough}
$$\sigma (B_{\parallel} ) - \sigma (0) =G_0
\ln\left[1+{\Delta_r(B_\parallel)\over B_\phi}\right].
$$
Using $B_\phi=0.06$~mT and $\Delta_r(B_\parallel)$ presented in
Fig.~\ref{f7} we obtain that the conductivity raising at
$B_\parallel=0.6$~T has to be about $1\,G_0$ | see the dotted line
in Fig.~\ref{f8}. In contrast to that, the measured longitudinal
magnetoconductance does not exhibit strong
$B_\parallel$-dependence. This points to importance of spin
effects in weak localization for our structure once again.

The expression for longitudinal magnetoresistance in the presence
of both roughness and Zeeman splitting can be obtained from
Eq.~(\ref{dsigma_tiltedRS}) in the limit $B_\perp\to 0$. As a
result we have explicitly
\begin{eqnarray}\label{inplaneMR}
&& \sigma (B_{\parallel} ) - \sigma (0) ={G_0\over
2}\Biggl[2\ln\left({B_\phi+B_s+\Delta_r\over
B_\phi+B_s}\right)\nonumber \\
&+&\ln\left({B_\phi+2B_s+\Delta_r\over B_\phi+2B_s}\right)
-\ln\left({B_\phi+\Delta_r+\Delta_s\over B_\phi}\right)\Biggr]\nonumber\\
&+& S(B_\phi/B_s +\Delta_r/B_s)-S(B_\phi/B_s),
\end{eqnarray}
where the function $S(x)$ is given by Eq.~(\ref{S}). Solid line in
Fig.~\ref{f8} is the longitudinal magnetoresistance calculated
according to Eq.~(\ref{inplaneMR}) with $\Delta_r(B_\parallel)$
and $\Delta_s(B_\parallel)$ parabolic dependences shown in
Fig.~\ref{f7} by solid lines. One can see satisfactory agreement
between experimental and calculated results. The dotted and dashed
line illustrate separate contributions of microroughness and
Zeeman effect in longitudinal magnetoconductance. One can see that
they are of opposite sign and compensate each other to a large
extent. It should be noted that such a compensation is not
universal. The sign and relative contribution of the considered
effects is determined by the parameters of concrete system
entering into Eq.~(\ref{inplaneMR}).

\section{Conclusion}
In conclusion, we have measured the anomalous magnetoresistance in
a field perpendicular, tilted and parallel to the QW plane.
Weak-antilocalization theory describes the data very well allowing
to obtain the spin and phase relaxation times. Our study
demonstrates that spin dephasing leading to weak antilocalization
is due to the Dresselhaus spin-orbit interaction which dominates
the Rashba term in our structures.

Application of an in-plane magnetic field is shown to destroy weak
antilocalization due to competition of Zeeman and microroughness
effects. From the analysis of longitudinal magnetoconductivity we
extracted the characteristics of short-range interface
imperfections and the value of the in-plane electron $g$-factor.

\section*{Acknowledgements} This work is financially supported by the
RFBR,  INTAS, CRDF, ``Dynasty'' Foundation --- ICFPM, and by the
Programmes of RAS and Russian Ministries of Science and Education.


\begin{thebibliography}{9}
\bibitem{slfcn}
The energy spectrum and electron charge distribution in the growth
direction have been obtained as a result of simultaneous solution
of the Schr\"{o}dinger and Poisson equations.
\bibitem{Lenya} In Ref.~\protect\onlinecite{ILP} the
expression for the weak-localization contribution to the
conductivity $\sigma_{WL}(B_{\perp})$ is given. We have calculated
the limit $\sigma_{WL}(B_{\perp} \to 0 )$ and arrived at the
Eq.~(\ref{dsigma_perp}).
\end{thebibliography}
\end{document}